\documentclass[12pt]{article}
\usepackage{graphicx}

\addtocounter{page}{0} \tolerance=5000

\newcommand{\beq}{\begin{equation}}
\newcommand{\eeq}{\end{equation}}
\newcommand{\beqn}{\begin{eqnarray}}
\newcommand{\eeqn}{\end{eqnarray}}

\newcommand{\lsim}{\mbox{$<$\hspace{-0.8em}\raisebox{-0.4em}{$\sim$}}}

\newcommand{\p}{\mathbf{p}}

\newcommand{\bI}{\mathbf{I}}

\newcommand{\br}{\mathbf{r}}

\newcommand{\si}{\mbox{{\boldmath$\sigma$}}}

\newcommand{\al}{\mbox{${\alpha}$}}

\newcommand{\De}{\mbox{${\Delta}$}}

\begin{document}
\begin{titlepage}
\begin{center}
{\Large \bf  Anapole moment of an exotic nucleus}
\end{center}

\vspace{1cm}

\begin{center}
K.E. Arinstein\footnote{k\_arin\_stein@ngs.ru}, V.F.
Dmitriev\footnote{dmitriev@inp.nsk.su}, and I.B.
Khriplovich\footnote{khriplovich@inp.nsk.su}
\\Budker
Institute of Nuclear Physics\\ 630090 Novosibirsk, Russia,\\ and
Novosibirsk University
\end{center}

\vspace{1cm}

\begin{abstract}
We consider the anapole moment of $^{11}$Be and demonstrate that
the contribution to it of the $1p_{1/2}$ level, which is
anomalously close to the ground state, is essentially compensated
for by the contribution of the continuum. Our estimate for this
anapole moment is $\kappa(^{11}{\rm Be})\simeq (0.07 - 0.08) g_n
$.
\end{abstract}

\vspace{1cm}

\end{titlepage}

{\bf 1.} The anapole moment is a special magnetic multipole
arising in a system which has no definite parity \cite{zel}. The
corresponding magnetic field looks like that created by a current
in toroidal winding.

For many years the anapole remained a theoretical curiosity only.
The situation has changed due to the studies of parity
nonconservation (PNC) in atoms. Since these tiny PNC effects
increase with the nuclear charge $Z$, all the experiments are
performed with heavy atoms. The main contribution to the effect is
independent of nuclear spin and caused by the parity-violating
weak interaction of electron and nucleon neutral currents.  This
interaction is proportional to the so-called weak nuclear charge
$Q$ which is numerically close (up to the sign) to the neutron
number $N$. Thus, in heavy atoms the nuclear-spin-independent weak
interaction is additionally enhanced by about two orders of
magnitude. Meanwhile, the nuclear-spin-dependent effects due to
neutral currents not only lack the mentioned coherent enhancement,
but are also strongly suppressed numerically in the electroweak
theory. Therefore, the observation of nuclear-spin-dependent PNC
phenomena in atoms had looked absolutely unrealistic.

However, it was demonstrated \cite{fk,fks} that these effects in
atoms are dominated not by the weak interaction of neutral
currents, but by the electromagnetic interaction of atomic
electrons with nuclear anapole moment (AM). Since the magnetic
field of an anapole, like that of a toroidal winding, is
completely confined inside the system, the electromagnetic
interaction of an electron with the nuclear AM occurs only as long
as the electron wave function penetrates the nucleus. In other
words, this electromagnetic interaction is as local as the weak
interaction itself, and in this sense they are indistinguishable.
The nuclear AM is induced by PNC nuclear forces and is therefore
proportional to the same Fermi constant $G=1.027\times 10^{-5}
m^{-2}$ (we use the units $\hbar=1, c=1$; $m$ is the proton mass),
which determines the magnitude of the weak interactions in general
and that of neutral currents in particular. The electron
interaction with the AM, being of the electromagnetic nature,
introduces an extra small factor into the effect discussed, the
fine-structure constant $\alpha=1/137$. Then, how it comes that
this effect is dominating?

The answer follows from the same picture of a toroidal winding. It
is only natural that the interaction discussed is proportional to
the magnetic flux through such a winding, and hence in our case is
proportional to the cross-section of the nucleus, i.e. to
$A^{2/3}$, where $A$ is the atomic number. Indeed, a simple-minded
model calculation leads to the following analytical result for the
dimensionless effective constant $\kappa$ which characterizes the
anapole interaction in the units of $G$ \cite{fks}:
\begin{equation}\label{an}
\kappa=\frac{9}{10}\,g\,\frac{\alpha \mu}{m r_0}\, A^{2/3}.
\end{equation}
Here $g$ is the effective constant of the P-odd interaction of the
outer nucleon with the nuclear core, $\mu$ is the magnetic moment
of the outer nucleon, $r_0=1.2$ fm. In heavy nuclei the
enhancement factor $A^{2/3}$ is close to 30 and compensates
essentially for the smallness of the fine-structure constant
$\alpha$. As a result, $\kappa$ is not so small in heavy atoms, it
is numerically close to 0.3.

The nuclear anapole moment was experimentally discovered in 1997
\cite{woo}. This result for the total effective constant of the
PNC nuclear-spin-dependent interaction in $^{133}$Cs is
\beq
\kappa_{tot}(^{133}{\rm Cs})=0.44(6).
\eeq
If one subtracts from this number the nuclear-spin-dependent
contribution of neutral currents, as well as the result of the
combined action of the ``weak'' charge $Q$ and the usual hyperfine
interaction, the answer for the anapole constant is
\beq\label{exp}
\kappa_{exp}(^{133}{\rm Cs})=0.37(6).
\eeq
Thus, the existence of an AM of the $^{133}$Cs nucleus is reliably
established.

The discussed result brings valuable information on PNC nuclear
forces. Of course, to this end it should be combined with reliable
nuclear calculations. The most detailed theoretical predictions
for this AM can be reasonably summarized, at the so-called ``best
values'' for the parameters of P-odd nuclear forces \cite{ddh}, as
follows \cite{dt1,hm}:
\beq\label{pre}
\kappa_{theor}(^{133}{\rm Cs})=0.15-0.21.
\eeq
There are good reasons to consider this prediction sufficiently
reliable, at the accepted values of the P-odd nuclear constants.

The comparison of the theoretical value (\ref{pre}) for the cesium
AM with the experimental result (\ref{exp}) indicates that the
``best values'' of \cite{ddh} somewhat underestimate the magnitude
of P-odd nuclear forces. In no way is this conclusion trivial. The
point is that the magnitude of parity-nonconserving effects found
in some nuclear experiments is much smaller than that following
from the ``best values'' (see review \cite{ah}). In all these
experiments, however, either the experimental accuracy is not high
enough, or the theoretical interpretation is not sufficiently
convincing. The experiment \cite{woo} looks much more reliable in
both respects. Still, further experimental investigations of
nuclear AMs are certainly of great interest.

\bigskip

{\bf 2.} In principle, the AM can be enhanced not only due to
large $A$, but also in the case when anomalously close to the
ground state of a nucleus there is an opposite-parity level of the
same angular momentum. In this connection, attention was attracted
in \cite{hu, hus} to exotic halo nuclei. In particular, the exotic
neutron-rich halo nucleus $^{11}$Be was considered therein. In
this nucleus the outer odd neutron is in the state $2s_{1/2}$, its
only bound excited level being $1p_{1/2}$ (the well-known
``inversion of levels''). The anomalously small energy separation
between these two levels of opposite parity,
\beq\label{int}
|\De E| = E(1p_{1/2}) - E(2s_{1/2}) = 0.32\, {\rm MeV},
\eeq
enhances by itself their P-odd mixing and thus the AM of this
nucleus. As pointed out in~\cite{hu,hus}, the small binding energy
of the odd neutron,
\beq\label{be}
|\De E_0| = 0.50\, {\rm MeV},
\eeq
affects the AM additionally, but in two opposite directions. On
one hand, it suppresses the overlap of the odd-neutron wave
function with the core, and thus suppresses the mixing of the
$2s_{1/2}$ and $1p_{1/2}$ levels due to the weak interaction
operator which looks as
\beq\label{w}
W=\,\frac{G}{\sqrt{2}}\,\frac{g_n}{2m}\,\{\si \p, \rho(r)\}\,;
\eeq
here $g_n$ is the effective constant of the P-odd interaction of
the outer neutron with the nuclear core, $\si$ and $\p$ are the
momentum and spin operators of the outer neutron, and $\rho(r)$ is
the spherically symmetric core density. On the other hand, the
small binding energy enhances the matrix element of $\br$ in the
anapole operator of the neutron
\beq\label{a}
\mathbf{a}=\,\frac{\pi e \mu_n}{m}\,\br \times \si\,;
\eeq
here $\mu_n=-1.91$ is the neutron magnetic moment.

The detailed calculation which takes into account the P-odd mixing
of the ground state with the $1p_{1/2}$ level only, results in the
following value for the effective anapole constant \cite{hus}:
\begin{equation}\label{b1}
\kappa_1(^{11}{\rm Be})= 0.17 g_n.
\end{equation}
Indeed, this value is 15 times larger than that given by the
estimate (\ref{an}) for $A=11$ (the neutron constant $g_n$ is
poorly known by itself, most probably $g_n\, \lsim \,1$).
Certainly, this enhancement of an AM in a light nucleus would be
of a serious interest, even if its possible experimental
implications are set aside.

However, such a strong enhancement of AM, as given in (\ref{b1}),
in a loosely bound nucleus does not look natural. In particular,
nothing of the kind happens in the deuteron. Even in the limit of
vanishing binding energy, when the energy interval between the
deuteron $s\,$ state and the continuum $p\,$ states tends to zero,
the deuteron AM in no way is enhanced~\cite{kk} (see also
\cite{sav}). As to the problem of $^{11}$Be discussed here, we
argue below that a strong cancellation between the contribution of
the bound $1p_{1/2}$ state (accounted for in (\ref{b1})) and that
of the continuum (omitted therein) takes place, resulting in a
serious suppression of the estimate (\ref{b1}).

\bigskip

{\bf 3.} We start with the general expression for the anapole
moment, as induced by operators~(\ref{w}) and (\ref{a}):
\beq\label{sum}
\langle 0|\mathbf{a}| 0 \rangle\,=\,\frac{G}{\sqrt{2}}\,\frac{\pi
e \mu g}{2m^2}\,\sum_n \frac{\langle 0 |\br \times \si | n \rangle
\langle n | \{\si \p, \rho(r)\} | 0\rangle\,+\, \langle 0|\{\si
\p, \rho(r)\} | n \rangle \langle n|\br \times \si|
0\rangle}{E(2s_{1/2})- E_n}
\eeq
To estimate the sum we use at first the closure approximation,
which is facilitated here by the same (negative) sign of all
energy denominators. After extracting some average value of
denominators, $-\bar{\De}$ ($\bar{\De} > 0$), and using the
completeness relation, the sum (\ref{sum}) reduces to
\beq\label{sum1}
\langle 0|\mathbf{a}| 0
\rangle\,=\,-\,\frac{G}{\sqrt{2}}\,\frac{\pi e \mu g}{2m^2
\bar{\De}}\, \langle 0 |\{[\br \times \si], \{\si \p, \rho(r)\}\}
| 0\rangle
\eeq
The thus arising effective operator transforms as follows:
\beq
\{[\br \times \si], \{\si \p, \rho(r)\}\}\,
=\,4(\mathbf{l}+\si)\,\rho(r)\,;
\eeq
here $\mathbf{l}$ is the orbital angular momentum of the valence
nucleon. (It is rather amusing that we arrive here at the same
combination $\mathbf{l}+\si$ which enters the expression for the
magnetic moment of a bound electron.)

In our case of $^{11}$Be, $\mathbf{l}=0$ and $\si=2\bI$, where
$\bI$ is the spin of the nucleus. Thus, here the expression for AM
reduces to
\beq\label{a1}
\langle 0|\mathbf{a}| 0 \rangle\,
=\,-\,\frac{G}{\sqrt{2}}\,\frac{4\pi e \mu_n g_n}{m^2 \bar{\De}}\,
\langle 0 |\rho(r)| 0\rangle \,\bI\,.
\eeq
With the standard prescription (see \cite{fks}) of deleting from
the expression for $\langle 0|\mathbf{a}| 0 \rangle$ the factors
$(G/\sqrt{2})\bI$ and multiplying the rest by $e
I(I+1)(-1)^{I+1/2-l}/(I+1/2)$, we arrive finally at the following
expression for the effective anapole constant:
\beq\label{b}
\kappa = \,\frac{3\pi \al \mu_n g_n}{m^2 \bar{\De}}\, \langle 0
|\rho(r)| 0\rangle \,.
\eeq

The expectation value $\langle 0 |\rho(r)| 0\rangle$ was
calculated by us with the same ground-state wave function
\[
R_{2s}(r)=\frac{2^{3/2}a^2[1-(r/a)^2]\exp(-r/r_0)}{r_0^{3/2}
\sqrt{45r^4_0+2a^4-12a^2r_0^2}}\,, \quad r_0=1.45\, {\rm
fm}\,,\quad a=2\, {\rm fm}\,,
\]
and core density
\[
\rho(r)=\rho_0\,\exp(-r^2/R_c^2)\,, \quad \rho_0=0.20\, {\rm
fm}^{-3}\,,\quad R_c=2 \,{\rm fm}\,,
\]
as those used in \cite{hus}. Thus obtained expectation value is
\beq\label{rho}
\langle 0 |\rho(r)| 0\rangle = 0.052\rho_0 = 0.01\, {\rm fm}^{-3}.
\eeq
Now, if $\bar{\De}$ is identified with the smallest energy
interval $E(1p_{1/2}) - E(2s_{1/2}) = 0.32\, {\rm MeV}$, the
numerical result is
\beq\label{b2}
\kappa(\bar{\De}= 0.32\, {\rm MeV})= -\, 0.036\, g_n.
\eeq

The comparison of (\ref{b2}) with (\ref{b1}) demonstrates that in
the last estimate the negative contribution of continuum states
overweighs the positive one of $1p_{1/2}$, with a small net result
which only slightly exceeds, if any, the typical value of $\kappa$
as given by (\ref{an}).

As expected, the small binding energy strongly suppresses $\langle
0 |\rho(r)| 0\rangle$ as compared to $\rho_0$ itself (see
(\ref{rho})). However, the expected enhancement of the matrix
element of $\br$ in the anapole operator (\ref{a}) is not
operative in (\ref{b2}) since on average this $\br$ is eaten up by
$\p$ in the weak interaction operator (\ref{w}). And the strong
suppression of $\langle 0 |\rho(r)| 0\rangle$ compensates in
(\ref{b2}) for the enhancement due to small energy intervals.

Estimate (\ref{b2}) can be improved considerably in the following
way. Its comparison with (\ref{b1}) demonstrates that with
$\bar{\De} = 0.32\, {\rm MeV}$ the contribution of the continuum
to $\kappa$ constitutes
\beq\label{cc}
\kappa_c(\bar{\De} = 0.32\, {\rm MeV})= -\, 0.036\, g_n
-\,0.17\,g_n =-\, 0.206\, g_n.
\eeq
With the continuum threshold at $\bar{\De}=|\De E_0| = 0.50\, {\rm
MeV}$, the continuum contribution is certainly overestimated by
(\ref{cc}). However, $\kappa_c$ can be easily recalculated for
more reasonable values of $\bar{\De}$ just by multipliyng
(\ref{cc}) by $0.32/\bar{\De}$.

Combining thus obtained improved values of $\kappa_c$ with
(\ref{b1}), we arrive at the following estimates for the anapole
moment of $^{11}$Be:
\begin{equation}\label{res}
\begin{array}{ccccc} \bar{\De},\, {\rm MeV}  & 0.6 & 0.7 &
0.8 \nonumber \\ &&&&\\ \kappa(^{11}{\rm Be})   & 0.060 g_n &
0.076 g_n & 0.088 g_n
\end{array}
\end{equation}

We believe that with all the uncertainties of our estimates
(\ref{res}) for the anapole moment of $^{11}$Be, they are more
reliable than (\ref{b1}). Most probably the real value of
$\kappa(^{11}{\rm Be})$ is around $(0.07 - 0.08) g_n$, i. e., it
is 2 - 3 times smaller than (\ref{b1}).

\end{document}